\documentclass[1p,fleqn]{elsarticle}
\usepackage{graphicx,hyperref,natbib}
\usepackage{amssymb,amsmath}

\begin{document}

\title{Propagation of waves from an arbitrary shaped surface -- a generalization of the Fresnel diffraction integral}
\author[lpi]{R.M. Feshchenko\corref{cor1}}
\ead{rusl@sci.lebedev.ru}
\author[lpi]{A.V. Vinogradov}
\ead{vinograd@sci.lebedev.ru}]
\author[lpi]{I.A. Artyukov}
\ead{iart@sci.lebedev.ru}
\cortext[cor1]{Corresponding author}
\address[lpi]{P.N. Lebedev Physical Institute of RAS, 53 Leninski Prospect, Moscow, 119991,  Russia}

\begin{abstract}\noindent
Using the method of Laplace transform the field amplitude in the paraxial approximation is found in the two-dimensional free space using initial values of the amplitude specified on an arbitrary shaped monotonic curve. The obtained amplitude depends on one {\it a priori} unknown function, which can be found from a Volterra first kind integral equation. In a special case of field amplitude specified on a concave parabolic curve the exact solution is derived. Both solutions can be used to study the light propagation from arbitrary surfaces including grazing incidence X-ray mirrors. They can find applications in the analysis of coherent imaging problems of X-ray optics, in phase retrieval algorithms as well as in inverse problems in the cases when the initial field amplitude is sought on a curved surface.
\end{abstract}
\maketitle
\section{Introduction}
Since the pioneering works of Leontovich and Fock \cite{fock1965electromagnetic} the parabolic wave equation (PWE) is widely used in many fields of physical and engineering sciences to describe the propagation of paraxial or quasi-paraxial beams in free space as well as in inhomogeneous media. It was successfully applied for solution of complex problems in laser physics \cite{sodha1976v}, electromagnetic radiation propagation \cite{levy2000parabolic}, underwater acoustics \cite{lee1995parabolic, spivack1994coherent}, X-ray optics \cite{sakdinawat2010nanoscale}, microscopy and lenseless imaging \cite{chapman2010coherent,thibault2010x}. This versatile nature of the PWE encourages searching for new applications and new methods of its solution. 

One of the areas that can benefit greatly from such new methods is the coherent X-ray imaging \cite{paganin2006coherent,nugent2010coherent}, which has been made possible by the development of powerful, versatile and coherent or quasi-coherent X-ray sources such as laboratory X-ray lasers \cite{suckewer2009x,ribic2012status, wang2008phase}, free electron lasers \cite{schmuser2008ultraviolet} and high order harmonics sources \cite{popmintchev2012bright}. The coherent X-ray imaging offers several advantages over traditional imaging techniques: a possibility of the lensless imaging and phase retrieval \cite{roy2011lensless}, diffraction imaging \cite{marathe2010coherent}, a sub-picosecond temporal resolution, etc. However, the coherent X-ray imaging, particularly in the reflective mode, poses a number of rather complicated mathematical problems \cite{fenter2008image,roy2011lensless}. One of them is appropriate description of the radiation field propagation starting from an arbitrary surface, which in general case can be non-flat, off-axis and tilted (see Figure 1). 

It should be noted that one of the underutilized mathematical properties of the PWE is a possibility to express the field amplitude in a part of free space through the initial values of amplitude specified on an arbitrary shaped line or surface. The case of a tilted straight line or plane was studied in \cite{artyukov2014optical, artyukov2014coherent, artyukov2016x, modregger2008fresnel} for applications in grazing incidence reflective microscopy and lithography. The present paper extends the PWE solution to a more general case of the initial surface being an arbitrary shaped one-dimensional monotonic curve. The parabolic initial curve and corresponding exact PWE solution will be also discussed.

\section{Direct problem on an arbitrary curve}

\subsection{General case}

\begin{figure}
\includegraphics[width=0.5\linewidth]{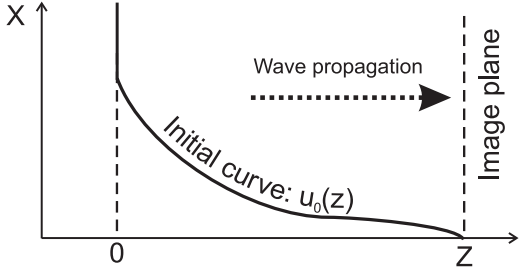}
\caption{A scheme showing the position of initial curve with field amplitude values $u_0(z)$ together with the image plane. The coordinate definitions are also shown.}
\label{f1}
\end{figure}

Let's consider the 2D PWE for the field amplitude $u$ in coordinates $(x,z)$\cite{levy2000parabolic}, where $z$ is the longitudinal coordinate along the beam propagation direction
\begin{equation}
i\frac{\partial u}{\partial z}+\frac{\partial^2 u}{\partial x^2}=0,
\label{1a}
\end{equation}
where it is assumed for simplicity that the wave number $k=1/2$. Let's assume that $u$ is known at some 2D initial curve (see Figure \ref{f1}), which is defined by the following equation
\begin{equation}
x-g(z)=0,\quad g(0)=0,\quad u_0(z)=u(g(z),z),
\label{1b}
\end{equation}
where $g(z)$ is a monotonic positive function. Let's introduce new coordinates $z'$ and $x'$ as
\begin{equation}
\begin{array}{rcl}
z'&=&z,\\
x'&=&x-g(z).
\end{array}\label{1c}
\end{equation}
In the new coordinates equation (\ref{1a}) can be rewritten as:
\begin{equation}
i\frac{\partial u}{\partial z}=i g'(z)\frac{\partial u}{\partial x}-\frac{\partial^2 u}{\partial x^2}
\label{1d}
\end{equation}
with the initial condition $u(0,z)=u_0(z)$ specified at the line $S'$ ($x'=0$) parallel to the axis $z$. In equation (\ref{1d}) prime marks of coordinates, for the sake of brevity, were omitted. The equation (\ref{1d}) can be solved using the Laplace transform by coordinate $x$
\begin{equation}
F(w,z)=\int\limits_0^\infty u(x,z)\exp(-wx)\;dx.
\label{1e}
\end{equation}
Applying it to equation (\ref{1d}) one can obtain the following differential equation for the function $F(w,z)$ 
\begin{equation}
i F'_z=i g'(z)wF-i g'(z)u_0-w^2F+wu_0+u_1,
\label{1f}
\end{equation}
where the transversal derivative
\begin{equation}
u_1=u'_x(0,z).
\label{1g}
\end{equation}
A solution of equation (\ref{1f}) can be written as
\begin{multline}
F(w,z)=-\int\limits_{-\infty}^z \left(g'(z')+i w\right)u_0(z')\exp\left[(i w+G)w(z-z')\right]\;dz'\\
-i\int\limits_{-\infty}^z u_1(z')\exp\left[(i w+G)w(z-z')\right]\;dz',
\label{1h}
\end{multline}
where
\begin{equation}
G(z,z')=\frac{\int\limits_{z'}^z g'(\xi)\;d\xi}{z-z'}=\frac{g(z)-g(z')}{z-z'}.
\label{1i}
\end{equation}
To obtain amplitude $u$ one should apply the reverse Laplace transform
\begin{equation}
u(x,z)=\frac{1}{2\pi i}\int\limits_{c-i\infty}^{c+i\infty}\exp[wx]F(w,z)\;dw,\quad c\ge0.
\label{1j}
\end{equation}
Since the expression under integral in (\ref{1j}) must not have any non-regularities in the right semi-plane of $w$ (including the imaginary axis), it is assumed that $c=0$. The absence of non-regularities is a direct consequence of a, so called, transparent boundary condition \cite{feshchenko2017exact}. Furthermore, taking into account that
\begin{equation}
I_1=\frac{1}{2\pi i}\int\limits_{-i\infty}^{+i\infty}\exp[wx+(i w+G)w(z-z')]\;dw=\frac{1}{2\sqrt{\pi i}\sqrt{z-z'}}\exp(i\Phi'),
\label{1k}
\end{equation}
and that
\begin{multline}
I_2=\frac{1}{2\pi i}\int\limits_{-i\infty}^{+i\infty}w\exp[wx+(i w+G)w(z-z')]\;dw=\\
\frac{1}{2\pi i}\frac{\partial}{\partial x}\int\limits_{-i\infty}^{+i\infty}\exp[wx+(i w+G)w(z-z')]\;dw=\frac{1}{2\sqrt{\pi i}\sqrt{z-z'}}\frac{\partial}{\partial x}\exp(i\Phi')=\\
\frac{i^{1/2}}{4\sqrt{\pi}\sqrt{z-z'}}\left(G+\frac{x}{z-z'}\right)\exp(i\Phi'),
\label{1l}
\end{multline}
where 
$$
\Phi'=\frac{(z-z')}{4}\left(G+\frac{x}{z-z'}\right)^2,
$$
and then substituting (\ref{1k}) and (\ref{1l}) into (\ref{1j}) and (\ref{1h}), one obtains the following final expression for the field amplitude $u$
\begin{multline}
u(x,z)=-\frac{1}{2\sqrt{-\pi i}}\int\limits_{-\infty}^z \frac{u_1(z')}{\sqrt{z-z'}}\exp(i\Phi')\;dz'+\\
\frac{1}{4\sqrt{\pi i}}\int\limits_{-\infty}^z (G-2g'(z'))\frac{u_0(z')}{\sqrt{z-z'}}\exp(i\Phi')\;dz'+\\
\frac{x}{4\sqrt{\pi i}}\int\limits_{-\infty}^z \frac{u_0(z')}{(z-z')^{3/2}}\exp(i\Phi')\;dz'.
\label{1m}
\end{multline}
Expression (\ref{1m}) depends on two functions -- $u_0$ and $u_1$ although the initial conditions (\ref{1b}) specify only one of them -- $u_0$. To find $u_1$ let's assume $x=0$ in formula (\ref{1m}). In this case, because
\begin{eqnarray}
&\lim_{x\to0}x\int\limits_{-\infty}^z \frac{u_0(z')}{(z-z')^{3/2}}\exp(i\Phi')\;dz'=\frac{2\sqrt{\pi}}{\sqrt{-i}}u_0(z),
\label{1n}
\end{eqnarray}
one can obtain the following integral equation of the Volterra first kind for $u_1$
\begin{multline}
u_0(z)=\\
-\frac{1}{\sqrt{-\pi i}}\int\limits_{-\infty}^z \frac{u_1(z')}{\sqrt{z-z'}}\exp(i\Phi)\;dz'+\frac{1}{2\sqrt{\pi i}}\int\limits_{-\infty}^z (G-2g'(z'))\frac{u_0(z')}{\sqrt{z-z'}}\exp(i\Phi)\;dz',
\label{1o}
\end{multline}
where
\begin{equation}
\Phi=\frac{(z-z')}{4}G^2,\label{1o1}
\end{equation}
which must be solved before the field amplitude $u$ can be calculated using formula (\ref{1m}).

A problem similar to one discussed here was considered in \cite{samarski1964} (see p. 521 -- boundary problems for regions with moving boundaries). The method used for PWE solution in the present work allowed one to obtain simpler expressions which have not been explicitly written before.

\subsection{Tilted line}

Let's consider the case when the initial curve (\ref{1b}) is a tilted (inclined) straight line. One can try to obtain the already known formula to check correctness of the derived result. So, one has
\begin{equation}
const=g'(z)=G(z,z')=-\tan\theta,
\label{1p}
\end{equation}
where $\theta$ is the angle between this tilted line and the axis $z$. Now it follows from (\ref{1o}) that $u_1$ can be expressed through $u_0$ as
\begin{equation}
u_1(z)=-\frac{i}{2}\tan\theta-\frac{i^{3/2}}{\sqrt{\pi}}\frac{\partial}{\partial z}\int\limits_{-\infty}^z\frac{u_0(\xi)}{\sqrt{z-\xi}}\exp\left[\frac{i}{4}\tan^2\theta(z-\xi)\right]\;d\xi.
\label{1q}
\end{equation}
Expression (\ref{1q}) is (as was said above) the transparent boundary condition for equation (\ref{1d}) \cite{feshchenko2017exact}. Now substituting (\ref{1q}) into (\ref{1m}) and applying the necessary transformations it is possible to show that 
\begin{equation}
u(x,z)=\frac{x}{2\sqrt{\pi i}}\int\limits_{-\infty}^z\frac{u_0(\xi)}{(z-\xi)^{3/2}}\times\exp\left[\frac{i(z-\xi)}{4}\left(\tan\theta-\frac{x}{z-\xi}\right)^2\right]\;d\xi.
\label{1r}
\end{equation}

Taking into account that $x$ in (\ref{1r}) is in reality $x'$ with omitted $'$, and substituting its expression $x'=x-z\tan\theta$ into (\ref{1r}) it can be obtained that
\begin{equation}
u(x,z)=\frac{x+z\tan\theta}{2\sqrt{\pi i}}\int\limits_{-\infty}^z\frac{u_0(\xi)}{(z-\xi)^{3/2}}\exp\left[\frac{i(x+\xi\tan\theta)^2}{4(z-\xi)}\right]\;d\xi.
\label{1r1}
\end{equation}
After changing the variable in the integral (\ref{1r1}) to $s=-\xi/\cos\theta$ one arrives at the following final expression for $u$
\begin{equation}
u(x,z)=\frac{x\cos\theta+z\sin\theta}{2\sqrt{\pi i}}\int\limits^{\infty}_{-z/\cos\theta}\frac{u_0(s)}{(z+s\cos\theta)^{3/2}}\times\exp\left[\frac{i(x+s\sin\theta)^2}{4(z+s\cos\theta)}\right]\;ds,
\label{1s}
\end{equation}
which coincides with equation (4) from \cite{artyukov2014optical} (taking into account that $k=1/2$). It can also be found in \cite{spivack1994coherent} where it is used to investigate the sound waves scattering from rough surfaces.

\section{Direct problem on a parabola}
Let's now assume that function $g(z)$ is the second order in $z$ i.e. it is a parabolic curve. Such a function $g(z)$ can in a general case can be written as
\begin{equation}
g(z)=az^2-bz,
\label{2a}
\end{equation}
where it is assumed that $a>0$ and $b\ge0$, that corresponds to a concave parabolic surface. To solve the PWE in this case one may use general formula (\ref{1m}) and (\ref{1o}). However it is simpler to take a different route. Let's introduce a new function $\varphi$ defined as
\begin{equation}
u(x,z)=\varphi(x,z)\exp\left[\frac{i}{2}(2az-b)x+\frac{i}{6a}(2az-b)^3\right].
\label{2b}
\end{equation}
Substituting now (\ref{2b}) into (\ref{1d}) one can obtain the following equation for $\varphi$
\begin{equation}
i\frac{\partial \varphi}{\partial z}=ax\varphi-\frac{\partial^2 \varphi}{\partial x^2},
\label{2c}
\end{equation}
where $x$ is actually $x'=x-g(z)$ as it was mentioned in the previous section. Now applying the Laplace transform (\ref{1e}) by $z\in[0,+\infty]$ one can obtain the following differential equation for image $F(p,t)$ of function $\varphi$
\begin{equation}
\frac{d^2 F}{d t^2}-tF=i a^{-2/3}\varphi_0,
\label{2d}
\end{equation}
where $t=(ax-i p)a^{-2/3}$ and $\varphi_0(x)=\varphi(x,0)$. It has the following solution satisfying the decaying boundary condition when $x\to+\infty$
\begin{multline}
F=C(p)Ai(e^{2\pi i/3}t)+\\
2\pi a^{-2/3}\left[Ai(t)\int\limits_{t}^{\infty}Ai(e^{2\pi i/3}t)\varphi_0(t')\,dt'+Ai(e^{2\pi i/3}t)\int\limits^{t}_{0}Ai\left(t\right)\varphi_0(t')\,dt'\right],
\label{2e}
\end{multline}
where $C(p)$ is an analytical function and $\mathop{Ai}(z)$ is the Airy function. Applying the reverse Laplace transform to (\ref{2e}) it is possible to obtain the following final expression for $\varphi$
\begin{multline}
\varphi(x,z)=\int\limits_{0}^{z}K(x, z-z')\varphi(0,z')\,dz'+\\
\frac{1}{2\sqrt{\pi i}z}\exp\left(-\frac{i z^3}{12}a^2\right)\left[\int\limits_{x}^{\infty}\exp(i\Phi_{+})\varphi_0(x')\,dx'+\int\limits_{0}^{x}\exp(i\Phi_{-})\varphi_0(x')\,dx'\right],
\label{2f}
\end{multline}
where the kernel in the first term, which is related to the value of the amplitude at the parabolic curve, and the initial condition $\varphi_0$ at $z=0$ are expressed as
\begin{eqnarray}
K(x, z)&=&\frac{1}{2\pi}\int\limits_{-\infty}^{+\infty}e^{i\xi z}\frac{Ai\left(e^{2\pi i/3}a^{-2/3}(\xi+ax)\right)}{Ai\left(e^{2\pi i/3}a^{-2/3}\xi\right)}\,d\xi,
\label{2g}\\
\varphi_0(x)&=&\exp\left[\frac{i}{2}bx+\frac{i b^3}{6a}\right]u(x,0),
\label{2h}
\end{eqnarray}
where
\begin{equation}
\Phi_{\pm}=\frac{(x-x')^2}{4z}\mp \frac{az}{2}(x-x').
\label{2h1}
\end{equation}
Finally, $u(x,t)$ can be obtained using formula (\ref{2b}). The expression for an infinite parabolic curve similar to (\ref{1m}) can be obtained from (\ref{2f}) by integrating in the first term from $-\infty$ instead of $0$ and omitting the second term. The final expression is
\begin{equation}
\varphi(x,z)=\int\limits_{-\infty}^{z}K(x, z-z')\varphi(0,z')\,dz',
\label{2i}
\end{equation}
which corresponds to the formula (\ref{1s}) in the linear case.

In a linear case when $a=0$ the second term in (\ref{2f}) turns into a single integral by $x$ from $0$ to $+\infty$ with a well known Fresnel type kernel while the integral in the second term turns into expression (\ref{1s}).

\section{Discussion and conclusion}

In the parabolic approximation the field amplitude in the free 2D space is expressed through its initial values on an arbitrary shaped monotonic curve. The obtained expression is a sum of two integrals dependent respectively on two functions defined on the initial curve: one is the field amplitude itself and the second one is the transversal derivative of the field amplitude. The transversal derivative shall be found from solution of a Volterra first kind integral equation, which arises naturally as a result of applying the transparent boundary condition to the general solution of 2D PWE. The obtained formula can be used in simulations of beam propagation in X-ray beamlines with non-flat grazing incidence mirrors, reflective X-ray imaging of tilted objects and generally of object having a complex shape, certification of super-polished surfaces, etc. It may open a new approach to inverse and phase retrieval problems for tilted objects with shaped or rough surfaces \cite{artyukov2014optical}.

In addition to the general solution for a curved line, an exact solution of the 2D PWE in the case when the initial curve is a finite convex parabola was independently derived. This solution depends only on the known amplitude on the initial parabolic curve (and on the semi-line at $z=0$) and is simpler to use than the general expression for an arbitrary initial curve discussed above. This solution can be used, for instance, to propagate the wave field amplitude from the surface of a parabolic mirror.

To summarize, the obtained solutions can advance the understanding of coherent imaging problems. Their main advantage over the solution for an inclined line follows from the fact that the real physical objects are rarely flat and therefore their imaging may not be easily simulated by using the formula with the initial condition specified on a line. This is especially true if the inverse or phase retrieval problems are considered.

\section*{Acknowledgments}
The work was supported by the Basic Research Programme of the Presidium of the Russian Academy of Sciences {\it Fundamental and applied problems of photonics and physics of new optical materials}.

\section*{References}
\bibliography{Curved_surf_OptCom} 
\bibliographystyle{elsarticle-num} 

\end{document}